\documentclass[%
twocolumn,%
showpacs,%
showkeys,%
preprintnumbers,%
amsmath,amssymb,%
page%
]{revtex4}
\usepackage{graphicx}
\usepackage{dcolumn}
\usepackage{bm}

\def\timeevol#1{\setbox1=\hbox{$\longmapsto$}\setbox2=\hbox{$
    \scriptstyle #1$}\copy1\kern-.5\wd1\kern-.5\wd2
    \raise-1.2\ht2\copy2\kern-.5\wd2\kern\wd1}

\newcommand{\be}{\begin{eqnarray}}
\newcommand{\ee}{\end{eqnarray}}

\newcommand{\ket}[1]{|#1\rangle}

\begin{document}


\title{Duality, Phase Structures and Dilemmas in Symmetric Quantum Games}
\author{
Tsubasa Ichikawa
and
Izumi Tsutsui
}

\affiliation{
High Energy Accelerator Research Organization (KEK),
Tsukuba, Ibaraki 305-0801, Japan
}

\date{February 22, 2006}

\begin{abstract}
Symmetric quantum games for 2-player, 2-qubit strategies are analyzed in detail by using a scheme in which all pure states in the 2-qubit Hilbert space are utilized for strategies.    We consider two different types of symmetric games exemplified by the familiar games, the Battle of the Sexes (BoS) and the Prisoners' Dilemma (PD).  These two types of symmetric games are shown to be related by a duality map, which ensures that they share common phase structures with respect to the equilibria of the strategies.   We find eight distinct phase structures possible for the symmetric games, which are determined by the classical payoff matrices from which the quantum games are defined.   We also discuss the possibility of resolving the dilemmas in the classical BoS, PD and the Stag Hunt (SH) game based on the phase structures obtained in the quantum games.    It is observed that quantization cannot resolve the dilemma fully for the BoS, while it generically can for the PD and SH if appropriate correlations for the strategies of the players are provided.
\end{abstract}

\pacs{02.50.Le, 03.67.-a, 87.23.Ge}
\keywords{quantum mechanics, game theory, entanglement}

\maketitle

\section{Introduction}
\setcounter{equation}{0}

Quantum game theory has attracted much attention in recent years as an interesting attempt to 
expand the scope of the conventional (classical) game theory, 
which is now a standard tool in various fields, most notably in
economics, for analyzing decision making processes.   The main thrust in the investigation of quantum game has come from the remarkable observation by Eisert {\it et al.}~\cite{EWL99} that the famous dilemma in the Prisoners' Dilemma (PD) game can be resolved if the players resort to strategies available in quantum theory.   
Subsequently, Marinatto and Weber \cite{MW00a} examined the dilemma in the Battle of the Sexes (BoS) game, 
another typical dilemma in game theory,  and observed that this, too, could be
resolved by adopting a quantum strategy involving a maximally entangled state.   
Application of quantum strategies to various other games, such as the Stag Hunt (SH) or the Samaritan's Dilemma game, has also been discussed in \cite{SOMI04}.   

These studies of the quantum games presented in \cite{EWL99, SOMI04} and \cite{MW00a} employ different schemes of quantum strategies,  
and it has turned out that the outcome of the analysis is highly dependent on the scheme used.   In fact, it has been pointed out in 
\cite{BH01, EW00, DXLZH02} that  in the scheme used in \cite{EWL99, SOMI04} the dilemma in PD can be resolved only if the strategic space is restricted artificially,  while a more recent study \cite{CT05} shows that there exists a new scheme in which the dilemma can be resolved even with a full strategic space.  Similarly, the resolution of the dilemma in the BoS has been argued using different reasonings depending on the schemes \cite{MW00b, NT04a, SOMI04} (for a generalized scheme, with no analysis on dilemmas, see \cite{NT04b}),  casting a doubt on the genuine nature of the resolution and, more importantly, the universality of the outcomes of quantum game in general.

The distinction among these schemes can be found in the definitions of quantum strategy, the strategic space which the players can exploit, and the way the quantum correlation (entanglement) is furnished.  These differences are crucial, because (as observed in \cite{BH01, EW00, DXLZH02}) different strategic spaces admit different stable solutions, and moreover the amount of entanglement required to resolve the dilemma will depend on the stage in the process it is measured.  Despite of this scheme-dependence, we have found in \cite{CT05} an intriguing phase structure for the quantum PD game, which is reminiscent of the \lq phase transition\rq\ of equilibrium solutions discovered earlier in \cite{DLXSWZH02} in a different scheme.   This suggests that the phase structures may exhibit a scheme-independent, intrinsic features of  quantum games under consideration.

The aim of the present paper is to support this idea by providing a convenient tool to analyze quantum games in general terms.   We consider 2-player, 2-qubit strategy games, which are the simplest nontrivial and yet have not been fully analyzed.  Using the scheme introduced in \cite{CT05}, we study in detail two types of \lq symmetric games\rq, exemplified by the BoS and PD games, respectively.   
We show that these two types of games are 
actually related by a duality map, which brings a game in one symmetric type into a game in the other symmetric type without changing the payoff in effect.   This is convenient because then we can use the outcome of the analysis of 
the BoS for the study of the PD, for instance.   
A quantum game may be regarded as a family of 
games provided by quantum correlations which are absent in classical settings,
and our geometric picture used to portray the correlation-family in this paper turns out to be quite convenient, especially for analyzing the phase structures of the game.   
We shall then see that symmetric games admit eight different types of phase structures with regard to the possible stable strategies (related to classical strategies) preferred by the players, and that these types are determined by the original classical games.   With these phase structures, we find that  the dilemma in the BoS cannot be resolved fully in our scheme, albeit alleviated to some extent \cite{MW00a}, irrespective of the amount of entanglement provided.  In contrast, the dilemma of the PD game can be resolved if a certain amount of correlations are introduced.   An analogous conclusion will also be drawn for the SH game, for which we find rather intricate phase structures for the full stable strategies compared to the BoS and PD games.

The plan of the paper is as follows.   We first introduce our scheme of quantum game in section II and present the duality map between the two types of symmetric games.  The phase structures of the symmetric games are studied in section III.  Section IV is devoted to the analysis of the BoS, PD and SH games, where we examine the resolution of the dilemmas based on the results obtained in section III.   Finally,  we give our conclusion and discussions in section V.

\section{Quantum Game and Duality for Symmetric Games}
\setcounter{equation}{0}

To begin with, we first recapitulate the classical 2-player, 2-strategy game and then introduce its quantum version following \cite{CT05}.
Let $i = 0, 1$, $j = 0, 1$ be the labels of the strategies available for the players, Alice and Bob, respectively, and let also $A_{ij}$ and  $B_{ij}$ be their payoffs when their joint strategy is $(i, j)$.  
In classical game theory, the game is said to be \lq symmetric\rq\ if $B_{ji} = A_{ij}$, that is,  if
their payoffs coincide when their strategies are swapped $(i, j) \to (j, i)$.   
To make a distinction from the other symmetry discussed shortly, we call such a game {\it $S$-symmetric} in this paper.   The PD and other familiar games such as the SH and the Chicken game (see, {\it e.g.}, \cite{FA03, SOMI04}) are $S$-symmetric games. 
Similarly, we call the game {\it $T$-symmetric} if $B_{1-j, 1-i} = A_{ij}$, that is, if the payoff matrices coincide
when the strategies of the two players are \lq twisted\rq\ as 
$(i, j) \to (1-j, 1-i)$.  
The BoS is an example of  $T$-symmetric games with the additional property $A_{01} = A_{10}$.
The payoffs in these $S$-symmetric and $T$-symmetric games are displayed
in the form of the bi-matrix $(A_{ij}, B_{ij})$ in Table \ref{Tdefault}.   

Given a payoff matrix, each player tries to maximize his/her payoff by choosing the best possible strategy, and if there exists 
a pair of strategies in which no player can bring him/her in a better position by deviating from it unilaterally, 
we call it a {\it Nash equilibrium} (NE) of the game.   The players will be happy if the NE is unique and fulfills certain conditions attached to the game ({\it e.g.}, Pareto-optimality or risk-dominance as mentioned later).   Even when there are more than one NE, the players  will still be satisfied if a particular NE can be selected over the other upon using some reasonings.  Otherwise, the players may face a dilemma, as they do in the case of the BoS and the PD.  

To introduce a quantum version of the classical game, we first regard Alice's strategies $i$ as vectors in a Hilbert space 
${\cal H}_A$ of a qubit.  Namely, corresponding to the classical strategies $i$  we consider vectors $\left| i \right>_A$ for $i = 0$ and 1 which are orthonormal in ${\cal H}_A$.   A general quantum strategy available for Alice is then represented by a normalized vector $\left| \alpha \right>_A$ (with the overall phase ignored, {\it i.e.}, a unit ray) in ${\cal H}_A$.  
Bob's strategy is similarly represented by a normalized vector $\left| \beta \right>_B$ in another qubit Hilbert space ${\cal H}_B$ spanned by orthonormal vectors
$\left| j \right>_B$ for $j = 0$ and 1 in ${\cal H}_B$.   
The strategies of the players can thus be expressed in the linear combinations,
\be
\begin{split}
\label{basis}
\left | \alpha \right>_A &=  \sum_i \xi_i(\alpha) \left | i \right>_A , \\
\left | \beta \right>_B &=     \sum_j \chi_j(\beta) \left | j \right>_B ,
\label{strgpl}
\end{split}
\ee
using the bases $\left| i \right>_A$,  $\left| j \right>_B$ which correspond to the classical strategies, 
with complex coefficients $\xi_i(\alpha)$, $\chi_j(\beta)$ which are functions of the parameters $\alpha$ and $\beta$
normalized as $\sum_i \vert \xi_i \vert^2 =  \sum_j \vert \chi_j \vert^2 = 1$.   The strategies of the individual players are, therefore, realized by local actions implemented by the players independently.

\begin{table}[t]
\begin{center}
$S$-symmetric: 
\setlength{\doublerulesep}{0.1pt}
\begin{tabular}{ c|c c }
\hline\hline
strategy & Bob 0 & Bob 1 \\
\hline
Alice 0 & $\,(A_{00} , A_{00})$ &$( A_{01} , A_{10})$ \\
Alice 1 & $\,(A_{10} ,A_{01})$ &$(A_{11} ,A_{11})$ \\
\hline\hline
\end{tabular}
\end{center}
\label{Sdefault}
\end{table}
\begin{table}[t]
\begin{center}
$T$-symmetric: 
\setlength{\doublerulesep}{0.1pt}
\begin{tabular}{c|cc}
\hline\hline
strategy & Bob 0 & Bob 1 \\
\hline
Alice 0 & $\,(A_{00} , A_{11})$ &$( A_{01} , A_{01})$\\
Alice 1&  $\,(A_{10} ,A_{10})$ & $(A_{11} ,A_{00})$\\
\hline\hline
\end{tabular}
\end{center}
\caption{Payoff bi-matrices $(A_{ij}, B_{ij})$ of the $S$-symmetric game (above) and the $T$-symmetric game (below).}
\label{Tdefault}
\end{table}

The {\it joint} strategy of the players, on the other hand,  is given by a  vector in the direct product Hilbert space $\mathcal{H} = \mathcal{H}_A \otimes \mathcal{H}_B$.  Here
lies one
of the crucial differences between the classical and quantum games: in quantum game theory, 
the joint strategy is specified not just by the choice of the strategies of the players but also by furnishing the quantum correlation (essentially the entanglement) between the individual strategies.  
Consequently,  the outcome of a quantum game rests also on a third party (or referee) that determines the correlation.   To be more explicit, using the product vector $| \alpha, \beta  \rangle = |  \alpha \rangle_A | \beta \rangle_B$ which is uniquely specified by the individual strategies, a vector   
in the total strategy space $\mathcal{H}$ is written as
\be
\left | \alpha, \beta; \gamma \right> 
= J(\gamma) \left | \alpha, \beta \right>
= J(\gamma) \left | \alpha \right>_A \left | \beta \right>_B,
\label{qstate}
\ee
where $J(\gamma)$ is a unitary operator providing
the quantum correlation between 
the individual strategies.  The  correlation factor $J(\gamma)$ with the parameter set $\gamma$ is designed to
exhaust all possible joint strategies available in ${\cal H}$.   
The payoffs for Alice and Bob are then given by the expectation values of some appropriate self-adjoint operators $A$ and $B$, respectively:
\be
\begin{split}
&\Pi_A(\alpha, \beta; \gamma) 
=   \left < \alpha, \beta; \gamma  | A  |  \alpha, \beta; \gamma  \right > ,\\ 
&\Pi_B(\alpha, \beta; \gamma) 
 =   \left < \alpha, \beta; \gamma  | B  |  \alpha, \beta; \gamma  \right > .
\label{payoff}
\end{split}
\ee
To sum up, a quantum game is defined formally by the triplet $\{{\cal H}, A, B\}$.

To choose the payoff operators $A$ and $B$, we require that, in the absence of quantum correlations
$J(\gamma) = I$ ($I$ is the identity operator in ${\cal H}$), the payoff values 
reduce to the classical ones when the players choose 
the \lq semiclassical (pure) strategies\rq\ $ \left| i, j \right> = |  i \rangle_A | j \rangle_B$, 
\be
\begin{split}
&\left< i', j' \right| A  \left| i, j \right>
=  A_{ij} \delta_{i' i}\delta_{j' j},
\\
&\left< i', j' \right| B  \left| i, j \right>
=    B_{ij} \delta_{i' i}\delta_{j' j}.  
\label{classicalpayoff}
\end{split}
\ee
Adopting, for simplicity,  the value $\gamma = 0$ for the reference point  at which $J(\gamma) = I$ holds, we find that,
for the uncorrelated product strategies $|  \alpha, \beta; 0  \rangle = | \alpha, \beta  \rangle$, 
the payoffs (\ref{payoff}) become
\be
\begin{split}
&\Pi_A(\alpha, \beta; 0) 
= \left < \alpha, \beta; 0  | A  |  \alpha, \beta; 0  \right >  = \sum_{i,j} x_i  A_{ij} y_j,
\\ 
&\Pi_B(\alpha, \beta; 0) 
= \left < \alpha, \beta; 0  | B  |  \alpha, \beta; 0 \right >  = \sum_{i,j} x_i  B_{ij} y_j,
\label{clpayoff}
\end{split}
\ee
where 
$x_i = \vert \xi_i\vert^2$, $y_j = \vert \chi_j\vert^2$ represent
the probability of realizing the strategies 
$ \left | i \right>_A$, $ \left | j \right>_B$ under the general choice $ \left | \alpha \right>_A$, $ \left | \beta \right>_B$ (see (\ref{strgpl})).    This ensures 
the existence of a {\it classical limit} at which
the quantum game reduces to the classical game defined by the payoff matrix $A_{ij}$, where now Alice and Bob are allowed to adopt
{\it mixed strategies} (see, {\it e.g.}, \cite{Rasmusen89}) with probability distributions
$x_i$, $y_j$ ($\sum x_i = \sum y_j = 1$) for strategies $i$, $j$.   We thus see that the quantum game is an extension of the classical game, 
in which the correlation parameter $\gamma$ plays a role similar to the Planck constant $\hbar$ in quantum physics in the technical sense that the classical limit is obtained by their vanishing limit.
Note that, since  the set $\{  \left| i, j \right>, \,  i, j = 0, 1\}$ forms a basis set in the entire Hilbert space ${\cal H}$, the payoff operators $A$ and $B$ are uniquely determined from the classical payoff matrices by (\ref{classicalpayoff}); in other words, our quantization is {\it unique}.

The aforementioned symmetries in classical game can also be incorporated into quantum game by using corresponding appropriate symmetry operators.  Indeed, by introducing the swap operator
\be
S | i, j  \rangle = | j, i  \rangle,
\ee
we see immediately that  in the classical limit the game is
$S$-symmetric, $\Pi_B(\beta, \alpha; 0) = \Pi_A(\alpha, \beta; 0) $, provided that the payoff operators $A$ and $B$ fulfill 
\be
B = S\, A\, S. 
\label{ssym}
\ee  
Analogously, if we introduce the notation $\bar i = 1 - i$ for $i = 0, 1$ ({\it i.e.},  $\bar 0 = 1$ and $\bar 1 = 0$) and thereby the twist operator,
\be
T  | i, j  \rangle  = | \bar j, \bar i  \rangle,
\ee
and the twisted states,
\be
\left | \bar\beta, \bar\alpha \right> :=  T  \left | \alpha, \beta \right> = \sum_{i, j} \xi_i(\alpha)\, \chi_j(\beta)\, | \bar j, \bar i  \rangle , 
\ee
we find that
in the classial limit the game is $T$-symmetric, 
$\Pi_B(\bar\beta, \bar\alpha; 0) = \Pi_A(\alpha, \beta; 0) $, provided that the operators fulfill 
\be
B = T\, A\, T.
\label{tsym}
\ee  

The symmetries can be elevated to the full quantum level if we adopt the correlation factor in the form \cite{CT05},
\be
J(\gamma) = e^{i \gamma_1S/2}e^{i \gamma_2T/2},
\label{corfac}
\ee
with real parameters $\gamma_i \in [0, 2\pi)$ for $i = 1, 2$  
\footnote{%
Apart from the irrelevant freedoms concerning the overall phase and the normalization, the dimensionality of the joint strategy space is 
$\dim{\cal H} - 2 = 6$ in real variables.   Since the individual strategies $\left | \alpha \right>_A$ and $ \left | \beta \right>_B$ are specified by $2 + 2 = 4$ parameters ({\it e.g.}, see (\ref{parstate})), the correlation factor must have another $2$ parameters to cover the full joint strategy space.   The actual construction of the correlation factor is far from unique, and our form (\ref{corfac}) is adopted  based on the convenience for the duality map.}.
In fact, one can readily confirm, using
$[S, T] = ST - TS = 0$, that under (\ref{tsym}) the game is $S$-symmetric 
\be
\Pi_B(\beta, \alpha; \gamma) = \Pi_A(\alpha, \beta; \gamma),
\label{spayoffrel}
\ee
even in the presence of the correlation (\ref{corfac}).   Similarly, the game is $T$-symmetric
\be
\Pi_B(\bar\beta, \bar\alpha; \gamma) = \Pi_A(\alpha, \beta; \gamma),
\label{tpayoffrel}
\ee
if (\ref{tsym}) is fulfilled.   Since the correlation parameters in $\gamma$ are arbitrary, the properties (\ref{spayoffrel}), (\ref{tpayoffrel}) imply that a symmetric quantum game
consists of a  ($\gamma$-parameter) family of games with the ($S$ or $T$) symmetry exhibited for each $\gamma$.

It is interesting to observe that these two types of symmetric games are actually related by unitary transformations.
To see this, let us introduce the operator $C_A$ which implements the conversion for Alice's strategies,
\be
C_A \ket{i, j} = \ket{\bar{i}, j}.
\label{cadualtr}
\ee
Note that $C_A$ satisfies
\be
C_A\, S\, C_A = T, \qquad
C_A\, T\, C_A = S.
\label{icv}
\ee
Consider then the transformation of strategy by unilateral conversion by Alice,
\be
\ket{\alpha, \beta; \gamma} \to C_A \ket{\alpha, \beta; \gamma}.
\label{convtr}
\ee
On account of the relation (\ref{cadualtr}) and the form of the correlation (\ref{corfac}), we find
\be
C_A \ket{\alpha, \beta; \gamma} =  \ket{\bar\alpha, \beta; \bar\gamma},
\label{convtr2}
\ee
with $\bar\gamma$ given by
\be
(\bar\gamma_1, \bar\gamma_2) = (\gamma_2, \gamma_1).
\label{convtgamma}
\ee

In addition, one may also
consider the transformation on the payoff operators,
\be
A \to \bar A = C_A\, A\,C_A,
\quad
B \to \bar B = C_A\, B\, C_A.
\label{eq:defbar}
\ee
One then observes that, if the game is $S$-symmetric fulfilling (\ref{ssym}),  the game defined by the transformed operators becomes $T$-symmetric,
\be
\bar B = T\, \bar A\, T.
\label{eq:s2t}
\ee
Analogously, if the game is $T$-symmetric fulfilling (\ref{tsym}), then the transformed operators define an $S$-symmetric game,
\be
\bar B = S\, \bar A\, S.
\ee
This shows that the conversion $C_A$ in (\ref{cadualtr}) provides a one-to-one correspondence, or {\it duality}, between an $S$-symmetric game and a $T$-symmetric game.   Some quantities in quantum game are invariant under the duality map while other are not.   For instance, 
the trace of the payoff,
\be
{\rm Tr}\,A = \sum_{i,j} A_{ij}= A_{00} + A_{01} +A_{10} +A_{11},
\ee
remains invariant
${\rm Tr}\,A \to {\rm Tr}\,\bar A = {\rm Tr}\,A$, whereas the alternate trace defined by
\be
\!\!\!\!\! \tau (A) = \sum_{i,j} (-)^{i+j}A_{ij}= A_{00} - A_{01} -A_{10} +A_{11},
\label{eq:trickone}
\ee
changes the sign
$\tau (A)  \to \tau (\bar A) = - \tau (A)$.

In formal terms, the two games given by $\{{\cal H}, A, B\}$ and $\{{\cal H}, \bar A, \bar B\}$ are dual
to each other in the sense that the payoff under the strategy
$ \ket{\alpha, \beta; \gamma}$ in one game is equivalent to the payoff under the dual strategy $C_A \ket{\alpha, \beta; \gamma} = \ket{\bar\alpha, \beta; \bar\gamma}$ in the other.  In particular, 
if the former game happens to be $S$-symmetric, then the latter is $T$-symmetric, and vice versa. 
This allows us to regard any two games as \lq identical\rq\ if their payoff operators are related by the duality map (\ref{eq:defbar}).  

Evidently,  the other conversion of the strategies by Bob 
$C_B \ket{i, j} = \ket{i, 1- j}$ can also be used to provide a similar but different duality.  Besides, their combination,
\be
C= C_A\otimes C_B,
\label{fulcon}
\ee
implements the renaming of the strategies $0 \leftrightarrow 1$ for both of the players, and yields a duality map which does not alter the type of symmetries of the game.   These duality maps $C_A$, $C_B$ and $C$ are used later to identify games defined from different classical payoff matrices.
We mention that these dualities are actually a special case of  the more general \lq gauge symmetry\rq\ in quantum game theory,  which is that 
the two games defined by
$\{{\cal H}, A, B\}$ and $\{{\cal H}, U A U^\dagger, U B U^\dagger\}$ with some unitary operator $U$
are dual to each other under the corresponding strategies
$ \ket{\alpha, \beta; \gamma}$ and $U \ket{\alpha, \beta; \gamma}$.   Thus the identification of games can be extended to those which are unitarily equivalent.

\section{Classification of $T$-symmetric Games}
\setcounter{equation}{0}

The foregoing argument suggests that in order to study the two types of symmetric games it is sufficient to consider either one of the two.  Moreover, even if the two games are of the same symmetric type, a further identification may be possible using the full conversion $C$.
In view of this, in the following we choose
the $T$-symmetric games and analyze the pattern of the allowed equilibria there.   To start with, we furnish the definition of an equilibrium which corresponds to the NE in classical game \footnote{In the present paper, we consider quantum joint strategies given by pure states only.  The space of pure states is not convex, and hence the Nash theorem \cite{N50}, which ensures the existence of NE for a classical game with mixed strategies, is no longer available.  The existence of QNE in quantum game is, therefore, non-trivial \cite{LJ03}.}.   A joint strategy $|\alpha^{\star},\beta^{\star}\rangle$ is called {\it quantum Nash equilibrium}  ({\it QNE}), if it satisfies
\be
\Pi_A(\alpha^{\star},\beta^{\star}; \gamma)\ge\Pi_A(\alpha,\beta^{\star}; \gamma),
\label{eq:QNEdefA}
\ee
for all $\alpha$, and also
\be
\Pi_B(\alpha^{\star},\beta^{\star}; \gamma)\ge\Pi_B(\alpha^{\star},\beta; \gamma),
\label{eq:QNEdefB}
\ee
for all $\beta$.   Note that the QNE is defined for a given $\gamma$ treated as a set of external parameters.  Below, we study the conditions for $\gamma$ under which a QNE appears.

To evaluate the payoffs explicitly, we write the strategies as
\be
\begin{split}
&|\alpha\rangle_A =  \cos{(\alpha_1/2)} |0\rangle_A + \sin{(\alpha_1/2)} \, e^{i\alpha_2} |1\rangle_A,
\\ 
&|\beta\rangle_B =  \cos{(\beta_1/2)} |0\rangle_B + \sin{(\beta_1/2)}\,  e^{i\beta_2} |1\rangle_B,
\label{parstate}
\end{split}
\ee
with angle parameters $\alpha_1, \beta_1 \in [0, \pi]$ and $\alpha_2, \beta_2 \in [0, 2\pi)$.  
For convenience, we henceforth adopt both of the ket notations  $|\alpha\rangle$ and $| i\rangle$ with the convention that 
$| 0\rangle$ and $| 1\rangle$ refer always to the latter notations.  Using (\ref{parstate})
we find that, for a $T$-symmetric game fulfilling (\ref{tsym}), the payoff for Alice reads
\be
\begin{split}
\Pi_A (\alpha,\beta;\gamma)
&= \frac{1}{4} \{ {\rm Tr}\, A +\tau (A)\cos{\alpha_1}\cos{\beta_1}\\
&+I_{+}^{\prime}(\gamma)\cos{\alpha_1} +I_{-}^{\prime}(\gamma) \cos{\beta_1}\\
&-I_{+}(\gamma) \sin{\alpha_1}\sin{\beta_1} \sin{\alpha_2}\cos{\beta_2}\\
&-I_{-}(\gamma) \sin{\alpha_1}\sin{\beta_1} \cos{\alpha_2}\sin{\beta_2}\}, 
\label{eq:payoff A}
\end{split}
\ee
where we have defined
\be
\begin{split}
I_{\pm}(\gamma)
&=G_+(\gamma)\pm G_-(\gamma),\\ 
I_{\pm}^{\prime}(\gamma)
&=G_+^{\prime}(\gamma)\pm G_-^{\prime}(\gamma),
\label{eq:trickone}
\end{split}
\ee
with
\be
\begin{split}
G_+(\gamma) 
&= (A_{00} - A_{11})\sin{\gamma_2 },\\ 
G_+^{\prime}(\gamma) 
&= (A_{00} - A_{11})\cos{\gamma_2 },\\ 
G_- (\gamma)
&= (A_{01} - A_{10})\sin{\gamma_1 },\\
G_-^{\prime}(\gamma) 
&= (A_{01} - A_{10})\cos{\gamma_1 }.
\label{eq:trick}
\end{split}
\ee
The payoff $\Pi_B (\alpha,\beta;\gamma)$ for Bob  is readily obtained from (\ref{eq:payoff A}) using the relation (\ref{tpayoffrel}).
The conditions for QNE (\ref{eq:QNEdefA}) and  (\ref{eq:QNEdefB}) imply 
\be
\begin{split}
\partial_{\alpha_i} \Pi_A (\alpha,\beta^{\star};\gamma)|_{\alpha^{\star}} &= 0 , \\
\partial_{\beta_i} \Pi_B (\alpha^{\star},\beta;\gamma)|_{\beta^{\star}} &= 0 , 
\label{eq:necc 2}
\end{split}
\ee
for $i = 1, 2$.  Besides, the Hessian matrices $\mathcal{P}_A$ and $\mathcal{P}_B$ given by
 \be
 \begin{split}
 \mathcal{P}_A (\alpha, \beta; \gamma)_{ij}&=\partial_{\alpha_i}\partial_{\alpha_j}\Pi_A (\alpha, \beta; \gamma), \\
  \mathcal{P}_B (\alpha, \beta; \gamma)_{ij}&=\partial_{\beta_i}\partial_{\beta_j}\Pi_B (\alpha, \beta; \gamma),
  \end{split}
     \label{eq:hescon}
\ee
must be both negative semi-deifinite,
\be  
\mathcal{P}_A (\alpha^\star, \beta^\star; \gamma)  \le 0,    
   \quad
\mathcal{P}_B (\alpha^\star, \beta^\star; \gamma)  \le 0.
   \label{eq:necc 3}
\ee   
Using (\ref{eq:payoff A}) we obtain, for example,
\begin{eqnarray}
& &\!\!\!\partial_{\alpha_2}\Pi_A(\alpha,\beta^\star; \gamma)|_{\alpha^\star} 
= - \partial_{\beta_2}\Pi_B (\alpha^\star, \beta; \gamma)|_{\beta^\star} \nonumber \\
& &\!\!\!= \frac{1}{4}\sin\alpha_1^\star\sin\beta_1^\star \, [ I_{-}(\gamma) \sin\alpha_2^\star\sin\beta_2^\star \nonumber \\
& &\qquad\qquad\qquad - I_{+}(\gamma) \cos\alpha_2^\star\cos\beta_2^\star].
\label{eq:necc A2new}
\end{eqnarray}
These conditions (\ref{eq:necc 2}) and (\ref{eq:necc 3}) will now be analyzed in detail.

\subsection{Edge strategies}

{}From (\ref{eq:necc A2new}) we see that an obvious set of solutions for (\ref{eq:necc 2}) are obtained if 
 \be
 \sin\alpha_1^\star=\sin\beta_1^\star=0.
 \label{eq:edge}
\ee
These have solutions given by the semiclassical pure strategies
$|i, j\rangle$ for $i, j = 0, 1$, {\it i.e.}, the four \lq edge\rq\ strategies,
\be
|0, 0\rangle, \quad
|1, 1\rangle, \quad
|0, 1\rangle, \quad
|1, 0\rangle,
 \label{eq:edgesol}
\ee
which correspond to classical pure strategies $(i, j)$.  Note, however, that these quantum edge strategies differ from the classical counterparts because the joint strategy is determined with the additional correlation factor $J(\gamma)$.   
Note also that on the edge strategies the unitary operation
$J(\gamma)$ yields only a one-parameter family of correlations for joint states $\left | i, j ; \gamma \right>$
in (\ref{qstate}), since one of the two factors in (\ref{corfac}) gives merely an overall phase.

 \begin{table}[t]
  \centering 
  \label{sufficient edge }
  \setlength{\doublerulesep}{0.1pt}
   \begin{tabular}{llll}   
\hline\hline
$|\alpha^*,\beta^*\rangle\quad$   &Hessian conditions\, & \qquad $\Pi_A(\alpha^\star, \beta^\star; \gamma)$\qquad\qquad\\
\hline
\, $|0,0\rangle$&$H_{+}>0, \, H_{-}>0$&$[{\rm Tr}\, A+\tau (A)+2G_+^{\prime}]/4$ \\
\, $|0,1\rangle$&$H_{-}<0$&$[{\rm Tr}\, A-\tau (A)+2G_-^{\prime}]/4$ \\
\, $|1,0\rangle$&$H_{+}<0$&$[{\rm Tr}\, A-\tau (A)-2G_-^{\prime}]/4$\\
\, $|1,1\rangle$&$H_{+}>0, \, H_{-}>0$&$[{\rm Tr}\,A+\tau (A)-2G_+^{\prime}]/4$\\
\hline\hline
\end{tabular}
  \caption{Hessian conditions and Alice's payoffs for edge strategies in $T$-symmetric games.  Bob's payoffs can be obtained from $\Pi_B(\alpha^*, \beta^*; \gamma) = \Pi_A(\bar\beta^*, \bar\alpha^*; \gamma)$.}
  \label{edgepayoffs}
\end{table}

\begin{figure*}[t]
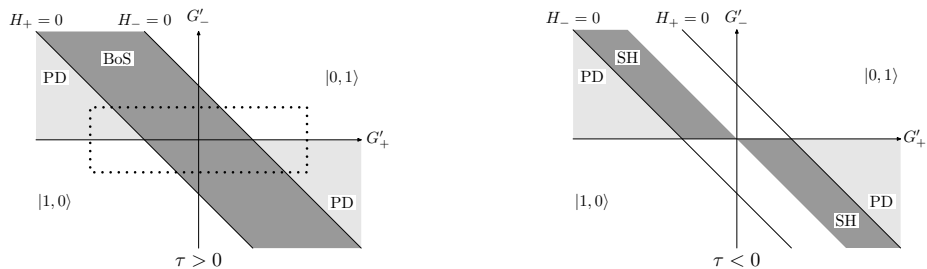

\begin{center}
\includegraphics[width=2in]{Classification.1}
\qquad\qquad\qquad
\includegraphics[width=2in]{Classification.2}
\caption{Phase structures of QNE in terms of edge strategies: $\tau (A)>0$ (above), $\tau (A)<0$ (below).   The names of the domains are borrowed from the classical games sharing the same characteristic dilemmas (see Table \ref{tab:class}).  Games in the domains without names are free from dilemmas within edge strategies and possess a single stable strategy $|1,0\rangle$ or $|0,1\rangle$ among at most two QNE.   The correlation family of a quantum game forms a rectangle on the plane, as shown by the dotted line for the case $\tau (A)>0$.}
\label{fig:taunegaedge}
\end{center}
\end{figure*}

{}For the edge states to become QNE, they also need to obey 
the Hessian conditions (\ref{eq:necc 3}), which pose different requirements for the states as
\be
\begin{split}
&|0, 0\rangle: \qquad H_{+}(\gamma)  > 0, \quad H_{-}(\gamma) > 0, \\
&|0, 1\rangle: \qquad H_{-}(\gamma)  < 0, \\
&|1, 0\rangle: \qquad H_{+}(\gamma)  < 0, \\
&|1, 1\rangle: \qquad H_{+}(\gamma)  > 0, \quad H_{-}(\gamma) > 0, 
\end{split}
\ee
where we have used 
\be
H_{\pm}(\gamma) =\tau (A)\pm I_+'(\gamma) ,
\ee
and ignored the cases of equalities for brevity.
These conditions and the payoffs for the edge solutions are summarized in Table \ref{edgepayoffs}.
To see when these conditions are fulfilled for
different $\gamma$, it is convenient to consider the plane coordinated by $(G_+^{\prime}, G_-^{\prime})$ with $G_\pm^{\prime}$ given in (\ref{eq:trick}).   One then sees that, as shown in Figure \ref{fig:taunegaedge}, the entire parameter region of $\gamma$ is mapped to 
a rectangular area in the centre of the $G_+^{\prime}$-$\,G_-^{\prime}$ plane with the horizontal length $L_{\rm h}$ and the vertical length $L_{\rm v}$ given by
\be
L_{\rm h} = 2|A_{00}-A_{11}|, \qquad L_{\rm v} = 2|A_{01}-A_{10}|.
\label{rectsize}
\ee

\begin{table}[t]
  \setlength{\doublerulesep}{0.1pt}
   \begin{tabular}{llll}   
\hline\hline
label \quad  &   \qquad QNE \qquad\quad\quad &   \qquad  characteristics \\
   \hline
BoS & $|0,0\rangle$ and $|1,1\rangle$  &  none \\  
PD  & $|1,0\rangle$ \makebox[.4mm] {} or \makebox[1mm] {}$|0,1\rangle$ & not Pareto optimal \\  
SH  & $|1,0\rangle$ and $|0,1\rangle$ & either payoff or risk dominant\\
\hline\hline
\end{tabular}
\caption{QNE and their characteristics in the domains on the $G_+^{\prime}$-$\,G_-^{\prime}$ plane classified by the labels of the classical games.  Both PD and SH games are mapped to their $T$-symmetric dual versions.
}
\label{tab:class}
\end{table}

It is worth noting that, at each of the midpoints of the four edges, the operation $J(\gamma)$ can yield a maximally entangled joint strategy state.   For instance, for $A_{01}>A_{10}$ the midpoint $(G_+',G_-')=(0,L_{\rm v}/2)$ corresponds to $J(\pi/2,0)$ under which the edge state $|01\rangle$ becomes $(|01\rangle + i|10\rangle)/\sqrt{2}$.   Similarly, for $A_{00}>A_{11}$ the midpoint $(G_+',G_-')=(L_{\rm h}/2, 0)$ corresponds to $J(0, \pi/2)$ under which the edge state $|00\rangle$ becomes $(|00\rangle + i|11\rangle)/\sqrt{2}$.   The four corners of the rectangle, on the other hand, correspond to $J(m\pi,n\pi)$ for $m, n = 0, 1$, which are $S$, $T$, $C$, and $I$ operations, and hence the resultant joint strategies are all unentangled.
On the $G_+^{\prime}$-$\,G_-^{\prime}$ plane, the Hessian conditions determine the domains of allowed edge QNE which are separated by the parallel lines $H_{\pm}(\gamma) = 0$ (see Figure \ref{fig:taunegaedge}).   Observe that the allowed edge QNE are different depending on the domains, and that the combinations of the QNE change when the sign of $\tau(A)$ is reversed.   Note that for
$\tau (A)>0$ all edge strategies in (\ref{eq:edgesol}) could arise as a QNE for some $\gamma$, whereas for $\tau (A)<0$ only $|0, 1\rangle$ and/or $|1, 0\rangle$ become QNE.

As will be seen shortly, as long as the edge strategies are concerned our quantum game is simulated by classical games possessing the corresponding NE.  
In view of this, we may characterize the domains on the $G_+^{\prime}$-$\,G_-^{\prime}$ plane by the
typical classical games sharing the same NE.  We do this by using the BoS, PD and SH as the representatives (see Table \ref{tab:class}).    Here, the label \lq BoS\rq\ is chosen to designate the domain of games possessing two edge QNE at $|0,0\rangle$ and $|1,1\rangle$, which is an obvious choice because the classical BoS game is $T$-symmetric and has the corresponding NE at $(i,j) = (0,0)$ and $(1,1)$.   None of these NE admits better payoffs to both of the players, simultaneously,  leading to the dilemma that they cannot decide on which strategy the should choose.
The domain \lq BoS\rq\ arises only for $\tau (A)>0$ and the required conditions are
\be
\hbox{BoS}: \qquad H_{+} <0, \quad H_{-}<0.
\ee
The domain fulfilling these forms a diagonal strip between the parallel lines $H_\pm = 0$ on the $G_+^{\prime}$-$\,G_-^{\prime}$ plane (see Figure \ref{fig:taunegaedge}).

To justify the assignment of the other labels, recall that the classical PD game is an $S$-symmetric game and has a NE at $(1,1)$ which is unique.  The problem of the game is that the NE is not {\it Pareto optimal}, {\it i.e.}, there exists another strategy which improves the payoffs for the two players, simultaneously, and this constitutes the dilemma. 
Upon quantization, the quantum PD, in the classical limit, will have one edge QNE at $|1,1\rangle$, which turns into $|0,1\rangle$ by the duality map  (\ref{cadualtr}) when it is employed to convert the PD into the $T$-symmetric dual version.   
For this reason,  we use \lq PD\rq\ to label the domain of those $T$-symmetric games possessing the edge QNE at $|0,1\rangle$ which is not Pareto optimal.   The Pareto optimality can be examined by comparing the payoff values with other strategies, and in the present case this is done essentially by comparing the payoffs between the two strategies $|1,0\rangle$ and $|0,1\rangle$.  From  Table \ref{edgepayoffs}, we see that this situation occurs when
\be
\hbox{PD}: \qquad H_{+}>0, \quad H_{-}<0, \quad G'_- < 0.
\ee
We also use the same label \lq PD\rq\ for the domain of games possessing a QNE at $|1,0\rangle$ which is not Pareto optimal, since those are identified by the full conversion $C$ in (\ref{fulcon}) with the standard quantum PD.    This is the case when we have
\be
\hbox{PD}: \qquad H_{+} <0, \quad H_{-}>0, \quad G'_- > 0.
\ee
As shown in Figure \ref{fig:taunegaedge}, the domains of PD appear both for $\tau (A)>0$ and $\tau (A)<0$.

The classical SH game, on the other hand, is an $S$-symmetric game which has two NE at $(0,0)$ and $(1,1)$, in which $(0,0)$ is {\it payoff dominant} ({\it i.e.}, better than $(1,1)$ in the payoff) and $(1,1)$ is {\it risk dominant} ({\it i.e.}, better than $(0,0)$ in the \lq average\rq\ over the choice of the other player).   The dilemma is that, while $(0,0)$ is Pareto optimal, $(1,1)$ is preferable for the minimal risk, which makes the players uncertain to decide which to choose.
Now, after the quantization and the application of the duality map to get the $T$-symmetric quantum version of the game, we will have two edge QNE  at $|1,0\rangle$ and $|0,1\rangle$ in the classical limit, with payoff dominant $|1,0\rangle$ and risk dominant $|0,1\rangle$.   We therefore use the label \lq SH\rq\ to name the domain in which the games possess the same QNE with the above property.   In the presence of correlations, we find  from Table \ref{edgepayoffs} that the payoff dominance of  $|1,0\rangle$ requires $G'_- < 0$.  The
risk dominance of $|0,1\rangle$ demands that the average payoff Alice receives under the choice $\vert 0\rangle_A$ be larger than that obtained under the choice $\vert 1\rangle_A$, which is ensured if $G'_+ + G'_- > 0$.
As in the case of the PD, the label \lq SH\rq\ is also used for the domain of games possessing the two QNE 
with payoff dominant $|0,1\rangle$ and risk dominant $|1,0\rangle$ for Alice, which are possible if 
$G'_- >0$ and $G'_+ + G'_- < 0$.
These domains \lq SH\rq\ are allowed only for $\tau (A)<0$ where $|1,0\rangle$ and $|0,1\rangle$ arise as QNE 
between the two parallel lines $H_\pm = 0$ on the $G_+^{\prime}$-$\,G_-^{\prime}$ plane.  Combined with the above additional conditions,  the SH domains are characterized by
\be
\hbox{SH}: \quad H_{+}>0, \,\, H_{-}>0, \,\, G'_- (G'_+ + G'_-) < 0.
\ee

As shown in Figure \ref{fig:taunegaedge}, the classification of the games leaves unlabeled domains on the $G_+^{\prime}$-$\,G_-^{\prime}$ plane for each of the cases $\tau (A)>0$ and $\tau (A)<0$.   For $\tau (A)>0$, we find two separate domains which contain games possessing a unique QNE, either at $|0,1\rangle$ or $|1,0\rangle$.  These QNE are Pareto optimal, and hence the games are free from the dilemma of the PD type.  For $\tau (A)<0$, we have two additional domains of games possessing QNE at $|0,1\rangle$ and $|1,0\rangle$, which are free from the dilemma of the SH.
This result suggests that, if the game under consideration can be driven to lie in these unlabeled domains by adjusting the correlations appropriately, then the original dilemma may be resolved under these correlations, at least within the realm of edge strategies.   In this respect, 
the phase diagram given by Figure \ref{fig:taunegaedge} provides a convenient basis to examine the problem of optimality of strategies in quantum games.  

\begin{figure}[t]
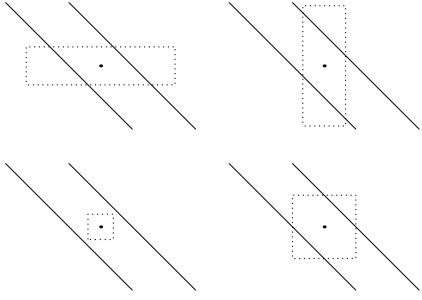

\includegraphics[width=1in]{framediagramv4.3}
\quad
\includegraphics[width=1in]{framediagramv4.4}
\caption{Four patterns of rectangles which are possible in relation to the parallel lines $H_{\pm} =0$ provide 
distinct phase structures for symmetric quantum games.  The rectangle may reduce to a line as we see in the case of the BoS later.}
\label{frame of game}
\end{figure}

Since the correlation-family of a   symmetric 
quantum game is mapped to a rectangle
on the $G_+^{\prime}$-$\,G_-^{\prime}$ plane,  we can classify quantum games in terms of the patterns of the rectangle formed on the plane.   As shown in Figure \ref{frame of game},  there are four types of rectangles,  determined from the values of $L_{\rm h}$ and $L_{\rm v}$ in (\ref{rectsize}), which are different in position with respect to the 
parallel lines $H_{\pm}(\gamma) = 0$ appearing in Figure \ref{fig:taunegaedge}.   Combining the two cases $\tau (A)>0$ and $\tau (A)<0$ which offer different structures for domains, we find that there are altogether eight classes of quantum games which have distinct phase structures of QNE in terms of edge strategies.

\begin{figure}[t]
\includegraphics[width=1in]{framediagramv4.1}
\quad
\includegraphics[width=1in]{framediagramv4.2}
\end{figure}

One of the advantages of the present quantization scheme is that it allows us to establish the connection between the classical and quantum games in a simple manner and thereby examine how \lq quantum\rq\ the game actually is.   To see this, let us
introduce the correlated payoff matrices
\be
\mathcal{A}(\gamma)=J^\dagger(\gamma)\, A\, J(\gamma), \quad
\mathcal{B}(\gamma)=J^\dagger(\gamma)\, B\, J(\gamma).
\ee
With these, the payoffs (\ref{payoff}) are expressed in terms of separable (uncorrelated) states
\be
\begin{split}
&\Pi_A(\alpha, \beta; \gamma) 
=   \left < \alpha, \beta  | \mathcal{A}(\gamma) |  \alpha, \beta  \right > ,\\ 
&\Pi_B(\alpha, \beta; \gamma) 
 =   \left < \alpha, \beta  | \mathcal{B}(\gamma)  |  \alpha, \beta  \right > .
\label{payoff2}
\end{split}
\ee
One may decompose each of the correlated payoffs  into \lq pseudo-classical\rq\ and \lq interference\rq\ terms as
\be
\mathcal{A}(\gamma)=\mathcal{A}^{\rm pc}(\gamma)+\mathcal{A}^{\rm in}(\gamma),
\ee
with
\be
\begin{split}
\mathcal{A}^{\rm pc}(\gamma)
&=\cos^2\frac{\gamma_1}{2} A+(\cos^2\frac{\gamma_2}{2} -\cos^2\frac{\gamma_1}{2})S\, A\,S\\
&\quad +\sin^2\frac{\gamma_2}{2} C\, A\,C,
\label{payoff2sup}
\end{split}
\ee
where $C$ is given in (\ref{fulcon}) and 
\be
\mathcal{A}^{\rm in}(\gamma)
=\frac{i}{2}\sin\gamma_1\left[A,S\right]+\frac{i}{2}\sin\gamma_2[A,T].
\ee
The point is that the pseudo-classical part $\mathcal{A}^{\rm pc}$ is diagonal and hence for separable strategies it can be interpreted as a classical payoff matrix.   In contrast, the interference part
$\mathcal{A}^{\rm pc}$ is non-diagonal and represents a non-classical contribution.   Accordingly, 
the payoff for Alice in a $T$-symmetric game is decomposed into the sum
$\Pi_A =\Pi^{\rm pc}_A +\Pi^{\rm in}_A$, where we have
\be
\Pi^{\rm in}_A(\alpha,\beta;\gamma) = 0,
\ee 
for edge strategies.  This observation confirms that our quantum game with edge strategies are, in effect, equivalent to the classical game with the payoff matrices $\mathcal{A}^{\rm pc}$ and $\mathcal{B}^{\rm pc}$ (the latter can be defined analogously for the correlated payoff $\mathcal{B}(\gamma)$).
Possible game theoretical interpretations of the pseudo-classical payoff  based on altruism and value-conversion have been noted in Ref.\cite{CT05}.

\subsection{Non-edge strategies}

To discuss QNE beyond the edge strategies (\ref{eq:edgesol}),
we recall (\ref{tpayoffrel}) and seek solutions which are
$T$-symmetric, $\ket{\alpha^*, \beta^*} = \ket{\bar\beta^*, \bar\alpha^*}$.  
In the representation (\ref{parstate}) of the state (which is defined up to an overall phase),  this translates into
\be
\alpha_1^\star-\beta_1^\star=\pi, \qquad
\alpha_2^\star+\beta_2^\star=\pi.
\label{ansatz}
\ee
Under the $T$-symmetric ansatz and the non-edge requirement $ \sin\alpha_1^\star\ne 0$,  the conditions in (\ref{eq:necc 2}) imply
\be
\begin{split}
\cos\alpha_1^\star \left[ \tau (A)-G_-(\gamma)\sin2\alpha_2^\star \right] - I_+'(\gamma) &= 0, \\
G_-(\gamma) \cos2\alpha_2^\star + G_+(\gamma) &= 0.
\label{fcond}
\end{split}
\ee
Besides, the Hessian condition (\ref{eq:necc 3}) implies
\be
G_-\sin2\alpha_2^\star \le 0.
\label{fcondtwo}
\ee

Before analyzing the solutions in detail, we observe that at the classical limit $\gamma = 0$ the above conditions are simplified to the single condition,
\be
\cos\alpha_1^\star = {{\sigma_+(A)}\over{ \tau (A)}},
\label{clsol}
\ee
with $\sigma_+(A)$ defined as
\be
\sigma_\pm(A) = (A_{00}-A_{11}) \pm (A_{01} - A_{10}).
\label{sigpm}
\ee
The condition (\ref{clsol}) has a solution when $\vert \tau(A) \vert \ge  \vert \sigma_+(A) \vert$, which is  equivalent to
\be
\begin{split}
&A_{00} \ge A_{10}  \quad\hbox{and} \quad A_{11} \ge A_{01}, \quad \hbox{or} \\
&A_{00} \le A_{10}  \quad\hbox{and} \quad A_{11} \le A_{01}.
\end{split}
\label{clnesol}
\ee
The solution for (\ref{clsol}) corresponds  to the NE for mixed strategies in classical games  with $x_1^* = y_1^* = \cos^2(\alpha_1^*/2)$, and (\ref{clnesol}) agrees precisely with the conditions for such nontrivial NE to arise.  It is important to note, however, that the non-edge QNE in quantum game and the mixed NE in classical game are completely different in the meaning of strategies.   Namely, the NE in classical game is relevant only for the situation where the games are repeated many times in which the players can consider probability distributions in choosing their strategies -- the mixed NE and pure NE belong to different categories conceptually.  In contrast,  the non-edge QNE in quantum game is a pure strategy and meaningful without repeating the game -- it belongs to the same category as the edge QNE.

\begin{figure}[t]
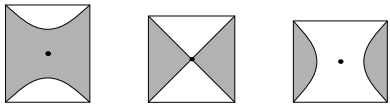

\includegraphics[width=.45in]{non-edge.1}
\qquad
\includegraphics[width=.46in]{non-edge.2}
\qquad
\includegraphics[width=.5in]{non-edge.3}
\caption{Possible patterns of allowed regions by (\ref{idelcon}) in the rectangle of the game under $L_{\rm h} <   L_{\rm v}$ (left),  $L_{\rm h} =   L_{\rm v}$ (middle) and $L_{\rm h} >   L_{\rm v}$ (right).  These regions are shaded, and the dot in the centre represents the origin of the $G_+'$-$G_-'$ plane.}
\label{non-edge}
\end{figure}

To discuss solutions for generic $\gamma$, we notice first that the second condition in (\ref{fcond}) determines $\alpha_2^*$, which can be used to determine $\alpha_1^*$ in the first condition.  
In terms of $\Delta(\gamma) := \sqrt{G_-^2 - G_+^2}$ for which we have
\be
\Delta^2 = (G'_+)^2 -  (G'_-)^2 - \sigma_+ \, \sigma_-,
\label{delcon}
\ee
the condition for the existence of $\alpha_2^*$ reads
\be
\Delta \ge 0.
\label{idelcon}
\ee
Notice that $\sigma_+ \, \sigma_- = (L_{\rm h}^2 - L_{\rm v}^2)/4$ measures the squared difference in length between the two edges of the rectangle of the game.  It follows that the regions allowed by (\ref{idelcon}) are those enclosed by the two hyperbolae $\Delta^2 = 0$ and the edges of the rectangle, which vary depending on the types of the rectangle (see Figure \ref{non-edge}).

On the other hand, by combining the two conditions in (\ref{fcond}) and (\ref{fcondtwo}) we see that 
the solution for $\alpha_1^*$ exists if 
\be
(H_+ + \Delta)(H_- + \Delta) \ge 0.
\label{hdelcon}
\ee
Using (\ref{delcon}),  one can readily depict  the regions where (\ref{hdelcon}) is fulfilled on the  $G_+^{\prime}$-$\,G_-^{\prime}$ plane.  The games belonging to the overlapped areas of  the above two regions admit non-edge QNE, and this is indeed possible if the payoff $A$ meets certain conditions, as illustrated by the SH game later.   When this happens,  the non-edge QNE, which we denote by $|\alpha^*,\beta^*; \gamma\rangle =|\alpha_{\rm ne},\beta_{\rm ne}; \gamma\rangle$, offers 
the same payoff (as ensured by the $T$-symmetry) for Alice and Bob,
\be
\begin{split}
&\Pi_A(\alpha_{\rm ne},\beta_{\rm ne};\gamma)
=\Pi_B(\alpha_{\rm ne},\beta_{\rm ne};\gamma)\\
&\quad =\frac{1}{4}
\left[{\rm Tr} \,A + \frac{\tau(A)\Delta (\gamma) - \sigma_+(A) \, \sigma_-(A)}{\tau (A)+\Delta(\gamma)}\right].
\label{eq:non-edge payoffs}
\end{split}
\ee
In particular, at the classical limit the payoff becomes
\be
\Pi_A(\alpha_{\rm ne},\beta_{\rm ne};0)
=
\frac{A_{00} A_{11} - A_{01}A_{10}}{A_{00} +  A_{11} - A_{01} - A_{10}},
\label{mixedpayoffs}
\ee
which is the familiar payoff expression for the mixed NE in classical $T$-symmetric games.  This shows that the non-edge QNE are actually an extension of the classical mixed NE.  In fact, at the classical limit, we find from $\Delta(0) = 0$ that the condition (\ref{idelcon}) is trivially fulfilled, and that (\ref{hdelcon}) reduces to
\be
\!\!\!\!\! H_+(0)\, H_-(0) = 4(A_{00} - A_{10})(A_{11} - A_{01}) \ge 0,
\label{clnecond}
\ee
which is exactly the condition for mixed NE (\ref{clnesol}).  In other words, if the classical game admits a mixed NE, then the quantum game defined from the classical game admits a QNE for a certain range of correlations including the classical limit.

To summarize, non-edge QNE may exist as an extension of mixed NE under various correlations in  quantum game theory, and their existence can be examined from the rectangle of the game 
specified from the
classical payoff matrix $A_{ij}$.   Game theoretical analysis, including the resolution of dilemma in classical game, should be made based on the combination of edge and non-edge QNE.

\section{Dilemmas in BoS, PD and SH}
\setcounter{equation}{0}

Having obtained the phase structures of symmetric quantum games for edge QNE as well as the conditions for non-edge QNE, 
we now examine if and how the typical dilemmas familiar in classical game theory -- the dilemmas in the BoS, the PD and the SH game -- can be  resolved in quantum game theory.    All of the dilemmas in these cases are intrinsically different, and there is no unique criterion for the resolution.    We thus consider the resolution based on the conventional requirements which are attached to the respective classical games, and find that the quantization of the games lead to considerably different outcomes for the three cases.

\subsection{Battle of the Sexes}

\begin{figure}[t]
\begin{center}
\includegraphics[width=1.5in]{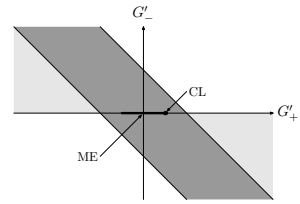}
\caption{Phase structure of edge QNE in the BoS game.  
The rectangle of the game is smashed to a line segment lying at the centre as shown by the dotted line, which is entirely contained in the BoS domain.  The right end point CL is the classical limit and the middle point ME represents the point where the maximally entangled correlation is realized.}
\label{BoS}
\end{center}
\end{figure}

The BoS game is  a special case of the $T$-symmetric game specified by the payoff matrix, 
\be                            
 A_{00}>A_{11}>A_{01}=A_{10}.
 \label{eq:constraint}
\ee                                
The degeneracy $A_{01}=A_{10}$ provides the $T$-symmetric game with an extra symmetry between the payoff matrices, that is,
\be
 B=T\, A\, T=C\, A\, C. 
\ee
On account of the degeneracy, we have $G_-(\gamma)=G_-^{\prime}(\gamma)= 0$, which implies that the parameter $\gamma_1$ drops out from our consideration of QNE.    Notice that the BoS defined by (\ref{eq:constraint}) has $\tau (A) > 0$ and that,  as shown in Figure \ref{BoS}, the rectangle of the game in the $G_+^{\prime}$-$\,G_-^{\prime}$ plane
is smashed to a line on the $G_+^{\prime}$-axis with length $L_{\rm h}$.   Notice also from
$A_{00}>A_{11}$ that the classical limit is found at the right end of the line.  
Now, an important point  to observe is that since
\be
\tau (A)-(A_{00}-A_{11})=2(A_{11}-A_{01})>0, 
\ee
the line segment of the game lies entirely within the BoS domain (see Figure \ref{BoS}).  
This shows that, so far as the edge strategies are concerned, even in the presence of the correlation $J(\gamma)$, the dilemma in the BoS does not disappear in quantum game.   Using (\ref{payoff2}) and (\ref{payoff2sup}), Alice finds her payoff $\Pi_A(i,i;\gamma)$ at the edge QNE $|\alpha^*,\beta^*\rangle = |i,i\rangle$ for $i = 0, 1$ as
\be
\Pi_A(i,i;\gamma)
= \cos^2\frac{\gamma_2}{2}\, A_{ii} + \sin^2\frac{\gamma_2}{2}\, A_{\bar{i}\bar{i}}.
\label{qneone}
\ee
Note that the correlation interpolates between the largest two payoff values $A_{00}$ and $A_{11}$, and hence
$\Pi_A(i,i;\gamma) \ge A_{11}$ for both of the edge QNE, $i = 0, 1$.

To see if the dilemma can be resolved by taking non-edge QNE into account, we first observe that for BoS the conditions (\ref{fcond}) are fulfilled for  $\gamma_2 = 0$, $\pi$ with arbitrary $\gamma_1$.   For that non-edge QNE, the payoff 
$\Pi_A(\alpha_{\rm ne},\beta_{\rm ne};\gamma)$  in (\ref{eq:non-edge payoffs}) reduces to (\ref{mixedpayoffs}) with $A_{01} = A_{10}$.  
At $\gamma_1 = 0$ this non-edge QNE  corresponds to the known mixed strategy NE in classical BoS, which cannot resolve the dilemma since the payoffs are strictly less than those obtained under the two edge QNE for both of the players.
The situation does not improve even for $\gamma_1 \ne 0$, because the payoffs are independent of  $\gamma_1$ for all strategies.    Moreover, on the general basis of the assignments (\ref{eq:constraint}) ({\it i.e.}, without making use of the ansatz (\ref{ansatz})),
one can confirm 
by looking at the Hessian condition (\ref{eq:necc 3}) 
that there is no non-edge QNE for BoS except for the one mentioned above.   Thus we find that under any correlations $\gamma$ for the non-edge QNE we have 
$\Pi_A(\alpha_{\rm ne},\beta_{\rm ne};\gamma) < A_{11}$ and hence 
\be
\Pi_A(i,i;\gamma) > \Pi_A(\alpha_{\rm ne},\beta_{\rm ne};\gamma), \quad \hbox{for} \quad i = 0, 1.
\ee

Although the dilemma does not disappear even in quantum BoS,  one may argue that the problem is somewhat mitigated at $\gamma_2 = \pi/2$ where the joint strategy state  is maximally entangled.  Indeed, under this correlation the payoffs for the two edge QNE (\ref{qneone}) for $i = 0, 1$ coincide and hence the choice of strategies becomes irrelevant for the players.  
The dilemma still remains in essence \cite{B00}, however, because the players, who cannot communicate, may inadvertently end up with a wrong strategy, $|0,1\rangle$ or $|1,0\rangle$, yielding the worst payoff $\Pi_A = \Pi_B =A_{01}$ (for all $\gamma$).   
A similar conclusion has been drawn for BoS in \cite{MW00a, MW00b} using a different quantization scheme with mixed quantum states, while a way out is suggested in an extended scheme \cite{NT04a}.    The analysis \cite{SOMI04} made in the scheme of  \cite{EWL99} yields a considerably different outcome, with infinitely many QNE with the payoffs lower than those of our edge QNE, indicating that the dilemma is unresolved unless some subtle reasoning (focal point effect) is invoked.

\subsection{Prisoners' Dilemma}

The PD game can also be analyzed in our scheme by converting it to a dual $T$-symmetric game using the map (\ref{eq:defbar}).   
The general $S$-symmetric PD in classical game theory may be defined by the payoff matrix for Alice $A_{ij}$ satisfying
\be
A_{10}>A_{00}>A_{11}>A_{01},
\label{eq:PDconst1}
\ee
together with Bob's payoff given by $B_{ij} = A_{ji}$.
Supplemental conditions (which is inessential for the following argument),
\be
2A_{00}>A_{01}+A_{10}>2A_{11},
\label{eq:PDconst2}
\ee
may also be imposed in order to render the strategies $(i,j) = (0,0)$ and $(1,1)$ the best and the worst of all possible strategies with respect to the sum of the payoffs \cite{Rasmusen89}.
The quantum PD is obtained by considering the self-adjoint operators $A$, $B$ fulfilling
(\ref{classicalpayoff}), and the duality map (\ref{eq:defbar}) yields the $T$-symmetric version of  the PD with the payoff operator $\bar A$ possessing the diagonal (classical) values
\be                            
(\bar{A}_{00}, \bar{A}_{01}, \bar{A}_{10}, \bar{A}_{11}) = (A_{10}, A_{11}, A_{00}, A_{01}).
\label{eq:dualofA}
\ee
In terms of the converted payoff values, the conditions (\ref{eq:PDconst1}) 
and (\ref{eq:PDconst2}) turn out to be
\be
\bar{A}_{00}>\bar{A}_{10}>\bar{A}_{01}>\bar{A}_{11},
\label{eq:const1bar}
\ee
and
\be
2\bar{A}_{10}>\bar{A}_{00}+\bar{A}_{11}> 2\bar{A}_{01}.
\label{eq:const2bar}
\ee
Note that under the duality map for strategies (\ref{convtr}) the parameters of the states (\ref{parstate}) acquire the change
\be
(\bar{\alpha}_1, \bar{\alpha}_2) = (\alpha_1+\pi, \, \pi - \alpha_2).
\label{eq:convtst}
\ee
In addition, the duality relation in the correlation (\ref{convtgamma}) amounts to 
$\gamma_1 \leftrightarrow \gamma_2$ in ${G}_{\pm}$ and ${G}'_{\pm}$.  To accommodate these changes caused by the duality map, we use notations such as 
\be
\bar{G}_{\pm} = {G}_{\pm}\vert_{A \to \bar A, \, \gamma \to \bar\gamma},
\quad
\bar{H}_{\pm} = {H}_{\pm}\vert_{A \to \bar A, \, \gamma \to \bar\gamma},
\ee
for our discussion of $T$-symmetric games.

\begin{figure}[t]
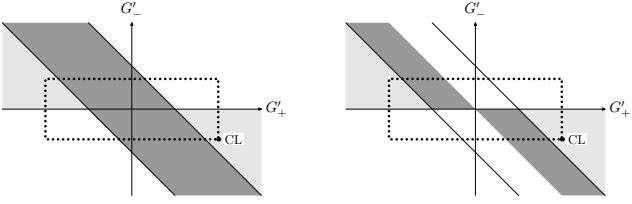

\begin{center}
\includegraphics[width=1.5in]{PD.1}
\qquad
\includegraphics[width=1.5in]{PD.2}
\caption{Phase structure of edge QNE in the ($T$-symmetrized) PD game 
for the cases $\tau (\bar{A}) > 0$ (left) and $\tau (\bar{A}) < 0$ (right).    For both of the cases, the rectangle 
of the game, whose edges are shown by dotted lines, extends to domains of no dilemmas.}
\label{PD}
\end{center}
\end{figure}

To examine the possible phase structures of the game, we observe  that
neither of the conditions (\ref{eq:const1bar}) and  (\ref{eq:const2bar}) determines the sign of $\tau (\bar{A})$. 
However, since (\ref{eq:const1bar}) implies that the classical limit $\gamma=0$ locates at the lower right corner of the rectangle of the game, the inequalities
\be
\begin{split}
\bar{H}_{+}(0) &=2(\bar{A}_{00}-\bar{A}_{10})>0,\\
\bar{H}_{-}(0) &=2(\bar{A}_{11}-\bar{A}_{10})<0, 
\label{pdineq}
\end{split}
\ee
obtained
at the classical limit $\gamma = 0$ from (\ref{eq:const1bar}) are sufficient to specify where the corner lies on the 
$G_+^{\prime}$-$\,G_-^{\prime}$ plane.  
The phase structures of the quantum PD game are then determined from the patterns of the rectangle in both of the cases $\tau (\bar{A}) > 0$ and $\tau (\bar{A}) < 0$, as illustrated in Figure \ref{PD}.   The outcome indicates that the correlation-family  given by the rectangle does extend to domains of no dilemmas.  
It follows that, as long as edge QNE are concerned, the quantum PD can be made dilemma-free when the correlations are furnished appropriately.

{}For a full resolution of the dilemma, we need to see whether a non-edge QNE, if any, alters our conclusion drawn from the edge QNE.  This can be examined from the analysis given in the previous section.  We then learn that, since the condition (\ref{clnesol}) is violated for (\ref{eq:const1bar}), there is no non-edge QNE at the classical limit.  We also realize that, for generic $\gamma$, the existence of non-edge QNE is dependent on the actual classical values of $A_{ij}$, and that for a wide range of payoff values centered at the standard ones
$(A_{10}, A_{00}, A_{11}, A_{01}) = (5, 3, 1, 0)$ used in the literature ({\it e.g.}, \cite{EWL99}), there exists no region fulfilling (\ref{idelcon}) and (\ref{hdelcon}) simultaneously, and hence no non-edge QNE.    Thus, our conclusion concerning the resolution of the dilemma does not change in these standard settings of the PD game.

\subsection{Stag Hunt}

The classical SH game is an $S$-symmetric game in which the payoff matrix for Alice fulfills the conditions,
\be
A_{00}>A_{10} \ge A_{11}>A_{01},
\label{eq:SHconst1}
\ee
which ensure that the strategies $(0,0)$ and $(1,1)$ are classical NE.   Among them, $(0,0)$ is payoff dominant while the other $(1,1)$ becomes risk dominant if 
\be
A_{10}+A_{11}>A_{00}+A_{01}.
\ee
Analogously to the PD, 
we quantize the SH according to (\ref{classicalpayoff}) and then $T$-symmetrize it by the duality map (\ref{eq:defbar}).  This yields the payoff operator $\bar A$ with the diagonal values (\ref{eq:dualofA}) obeying
\be
\bar{A}_{10}>\bar{A}_{00} \ge \bar{A}_{01}>\bar{A}_{11},
\label{eq:SHconst1bar}
\ee
and
\be
\bar A_{00}+\bar A_{01}>\bar A_{10}+\bar A_{11}.
\label{eq:SHconst2bar}
\ee

\begin{figure}[t]
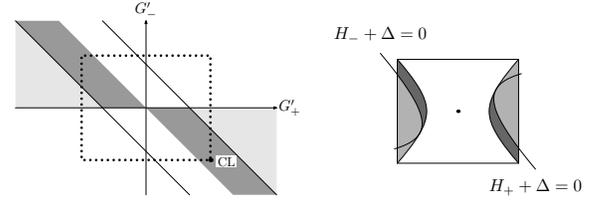

\begin{center}
\includegraphics[width=1.5in]{SHedge.1}
\quad
\includegraphics[width=1.3in]{SHnonedge.1}
\caption{Phase structures of edge QNE (left) and non-edge QNE (right) in the ($T$-symmetrized) SH game.  
{}For edge QNE, the rectangle of the game extends to domains of no dilemmas.  
{}For non-edge QNE, the allowed regions by (\ref{idelcon}) are of the third type in Figure \ref{non-edge}, and the two narrow regions overlapped with (\ref{hdelcon}) shown in thick gray indicate the domains where 
a non-edge QNE appears. }
\label{SH}
\end{center}
\end{figure}

Note that (\ref{eq:SHconst1bar}) implies $\tau (\bar{A}) < 0$.  It also shows that
the classical limit is at the lower right corner of the rectangle of the game, and that we have 
$\bar{H}_{\pm}(0) < 0$.    From this we can determine the position of the rectangle on the 
$G_+^{\prime}$-$\,G_-^{\prime}$ plane as shown in Figure \ref{SH}.   
The phase structure of the quantum SH game then suggests that, as in PD, the correlation-family  given by the rectangle extends to domains without dilemmas.   Within the edge strategies, the dilemma of the SH can therefore be resolved in quantum game, if one adjusts the correlations appropriately.  The payoffs $\Pi_A(i,\bar{i};\gamma)$ at the edge QNE 
$|\alpha^*,\beta^*\rangle = |i,\bar{i}\rangle$ for $i = 0, 1$ read
\be
\Pi_A(i,\bar{i};\gamma)
= \cos^2\frac{\gamma_1}{2}\, \bar A_{i\bar{i}} + \sin^2\frac{\gamma_1}{2}\, \bar A_{\bar{i}i},
\label{qneone}
\ee
which fall within the range of the payoffs of the two classical NE,
$\bar A_{10} \ge \Pi_A(i,\bar{i};\gamma) \ge \bar A_{01}$.

The classical SH game admits a mixed NE, and accordingly the quantum SH admits a non-edge QNE for a range of correlations including the classical limit, as can be confirmed explicitly by examining the condition (\ref{clnecond}).   To see where such correlations occur on the $G_+^{\prime}$-$\,G_-^{\prime}$ plane, we consider the lines of equality $H_\pm + \Delta = 0$ determined by the condition (\ref{hdelcon}), which are rewritten as 
\be
G_+^{\prime} = - \, G_-^{\prime} - {{\tau^2 + \sigma_+\sigma_-}\over{2(G_-^{\prime} \pm \tau)}}.
\label{neline}
\ee
{}For the SH we find $\tau^2 + \sigma_+\sigma_- > 0$ from (\ref{eq:SHconst1bar}) and (\ref{eq:SHconst2bar}).  The domains where the non-edge QNE arise are then found to be surrounded by the hyperbolae $\Delta = 0$ and the curves (\ref{neline}), both of which come in contact at
\be
(G_+^{\prime}, G_-^{\prime}) 
= 
\mp {1\over{2\tau}}\left(\tau^2 + \sigma_+\sigma_-, \tau^2 - \sigma_+\sigma_- \right).
\ee
As illustrated in Figure \ref{SH}, these domains are given by two narrow regions along the left and right edges of the rectangle of the game, indicating that under generic correlations the non-edge QNE does not spoil the resolution of the dilemma in terms of edge QNE.  
It is, however, conceivable that the non-edge QNE, in the region where
it is allowed, could alter the nature of the dilemmas, that is, the non-edge QNE could be both payoff and risk dominant under some particular correlations in the domains of SH, or it could pose a new dilemma in the domains where there was no dilemma originally.  These possibilities should be examined for the actual values of the payoff matrix (\ref{eq:SHconst1bar}), but the analysis with the standard values $({A}_{00},{A}_{10},{A}_{11}, {A}_{01}) = (4, 3, 3, 0)$ given in Table \ref{SHME} suggests that these are not likely to occur unless the payoff values are fine-tuned.

\begin{table}[t]
\begin{center}
\hspace{-.68cm} CL: 
\setlength{\doublerulesep}{0.1pt}
\begin{tabular}{ c|c c c }
\hline\hline
strategy & Bob $|0\rangle$ & Bob $|1\rangle$ & Bob $|\beta_{\rm ne}\rangle$ \\
\hline
\hspace{-.29cm}Alice $|0\rangle$ & $\,(3,0)$ &$(3,3)$&$(3,3/4)$ \\
\hspace{-.29cm}Alice $|1\rangle$ & $\,(4,4)$ &$(0,3)$&$(3,15/4)$ \\
Alice $|\alpha_{\rm ne}\rangle$ & $\,(15/4,3)$ &$(3/4,3)$&$(3,3)$ \\
\hline\hline
\end{tabular}
\end{center}
\label{SHCL}
\end{table}

\begin{table}[t]
\begin{center}
ME:
\setlength{\doublerulesep}{0.1pt}
\begin{tabular}{c|ccc}
\hline\hline
strategy & Bob $|0\rangle$ & Bob $|1\rangle$ & Bob $|\beta_{\rm ne}\rangle$ \\
\hline
\hspace{-.29cm}Alice $|0\rangle$ & $\,(3 , 0)$ &$( 7/2 , 7/2)$&$(3,0)$\\
\hspace{-.29cm}Alice $|1\rangle$ &  $\,(7/2 , 7/2)$ & $(0,3)$&$(7/2 , 7/2)$\\
Alice $|\alpha_{\rm ne}\rangle$ &  $\,(7/2 , 7/2)$ & $(0,3)$&$(7/2 , 7/2)$\\
\hline\hline
\end{tabular}
\end{center}
\caption{Quantum payoff bi-matrices $(\Pi_A, \Pi_B)$ of the SH game for edge QNE and non-edge QNE.  We used the values  $\bar{A}_{10} = 4$, $\bar{A}_{00} = \bar{A}_{01}=3$ and $\bar{A}_{11} = 0$ (obtained from those mentioned in the text) to evaluate the payoffs at the classical limit CL and the maximally entangled point ME, which are given by  $(G_+^{\prime},G_-^{\prime})=(3,-1)$ and $(3,0)$, respectively.
The presence of the non-edge QNE worsens the risk balance between the two edge QNE as we increase the amount of entanglement, but the dilemma disappears at ME where their payoffs become identical, for which the non-edge QNE does not contribute.}
\label{SHME}
\end{table}

\section{Conclusion and Discussions}

In this paper, we studied the phase structures of symmetric quantum games with respect to the stable strategies (QNE) available by pure states in quantum mechanics.   For quantization of  classical games we adopted the scheme \cite{CT05} which defines a unique correlation-family of quantum games from a classical game,  allowing for all possible strategies 
realized by pure states, entangled or not.    The correlation-family is projected onto a rectangular area 
in the $G_+^{\prime}$-$\,G_-^{\prime}$ plane, where the phase structures of both the edge and non-edge QNE in the game can readily be recognized.
We have found that for symmetric games there arise altogether eight different classes of phase structures for edge QNE depending on the payoff matrices of the classical game we started with.  This result gives a more detailed account of the phase structures mentioned in \cite{EWL99} and discussed later in \cite{DXLZH02, DLXSWZH02}.

The symmetric games considered in this paper consist of two types, $T$-symmetric and $S$-symmetric.
We have presented a unified framework to treat
them by means of a duality map, which enables us to use the results of the analysis of $T$-symmetric games for studying $S$-symmetric games and vice versa.  
As an example of the $T$-symmetric game, we studied the BoS which is known to be afflicted with a dilemma classically.  We have found that the dilemma in the BoS cannot be resolved fully  (albeit it can be alleviated) with strategies given by pure states, even if we go over to quantum game where arbitrarily entangled states are utilized.   Thus, the previous observation made in \cite{MW00a, MW00b} remains essentially unchanged even in our enlarged scheme of quantum game, while the outcome is considerably different from those obtained in other schemes \cite{SOMI04, NT04a}.   As for the $S$-symmetric game, we examined the PD and the SH to observe that for both of the games the correlation-family contains a phase which is free from dilemmas under edge QNE.  Since the standard PD does not admit non-edge QNE, we concluded that  for the PD the classical dilemma disappears after quantization.  For the SH, on the other hand, there exists a non-edge QNE which does not affect the resolution realized by the edge QNE, generically.
In short, quantum entanglement can resolve classical dilemmas for certain games, and the games for which this is possible can be judged from the classical payoff matrices.   
We remark that entanglement is necessary for the resolution of the dilemmas in our scheme, and that this is so in any other schemes of quantum games in which the resolution is possible and the classical games are recovered in the limit where the joint strategies become separable as in (\ref{clpayoff}).  However,  the actual amount of entanglement required depends on the scheme used (because the class of families considered may be scheme-dependent) as well as  on the values of the classical payoffs.

Compared to most other schemes proposed so far, our scheme of quantum game is distinguished in the specification of strategies and correlations which are expressed in the ordering of operations implementing them.   Namely, in our scheme
the players first make their choice of strategies independently, by performing the corresponding local unitary 
transformations on a fixed separable state, before a third party furnishes a correlation for the local states.   The player's strategy is represented by a quantum state, not by the local unitary transformation as considered in 
\cite{EWL99}.  The advantage for this is that different pure states used to specify the strategies yield different outcomes of the payoff in general, while this is not ensured if unitary transformations are regarded as strategies.  In fact, it has been pointed out \cite{BH01} that unitary transformations become redundant ({\it i.e.}, different unitary operations give the same quantum states) when the strategies are maximally entangled.   Obviously, the choice of quantization scheme is directly related to the question of the role of the third party which provides the quantum correlation in the game, and this has not been fully explored yet.  
In this regard,  we have found here a number of interesting features of quantum games which are commonly observed in various different schemes, in the phase structures of the QNE and the resolution of dilemmas in some of the familiar games.    We hope that these findings will help uncover the core elements  -- independent of the scheme employed -- in quantum games, which are crucial for laying
a solid foundation of quantum game theory.

\begin{acknowledgments}
We thank T. Cheon for helpful discussions.
This work is supported by the Grant-in-Aid for Scientific Research, 
No.~13135206 and No.~16540354, of the Japanese Ministry of Education, 
Science, Sports and Culture.
\end{acknowledgments}


  \end{document}